# USING SPECTRAL RADIUS RATIO FOR NODE DEGREE TO ANALYZE THE EVOLUTION OF SCALE-FREE NETWORKS AND SMALL-WORLD NETWORKS

Natarajan Meghanathan

Jackson State University, 1400 Lynch St, Jackson, MS, USA
natarajan.meghanathan@jsums.edu

## ABSTRACT

*In this paper, we show the evaluation of the spectral radius for node degree as the basis to analyze the variation in the node degrees during the evolution of scale-free networks and small-world networks. Spectral radius is the principal eigenvalue of the adjacency matrix of a network graph and spectral radius ratio for node degree is the ratio of the spectral radius and the average node degree. We observe a very high positive correlation between the spectral radius ratio for node degree and the coefficient of variation of node degree (ratio of the standard deviation of node degree and average node degree). We show how the spectral radius ratio for node degree can be used as the basis to tune the operating parameters of the evolution models for scale-free networks and small-world networks as well as evaluate the impact of the number of links added per node introduced during the evolution of a scale-free network and evaluate the impact of the probability of rewiring during the evolution of a small-world network from a regular network.*

## 1. INTRODUCTION

Network analysis and visualization of large complex real-world networks, ranging anywhere from social networks [1][2], co-authorship networks [3], Internet [4], World wide web [4] to biological networks [5] and etc is an actively researched area in recent years. The strength of network analysis is to abstract the complex relationships between the members of the system in the form of a graph with nodes (comprising of the constituent members) and edges (weighted or unit-weight as well as directed or undirected, depending on the nature of the interactions) and study the characteristics of the graph with respect to one or more metrics (like node degree, diameter, clustering coefficient and etc). The adjacency matrix A(G) of the network graph G essentially captures the presence of edges between any two vertices. For any two vertices $i$ and $j$ in graph G, the entry in the $i^{th}$ row and $j^{th}$ column of A(G) is 1 if there is an edge from vertex $i$ to vertex $j$ and 0 otherwise. This paper focuses on undirected graphs (edge $i$-$j$ exists in both the directions: $i$-->$j$ as well as from $j$-->$i$) and node degree as the metric under study.

Spectral decomposition is a method of projecting the characteristics of a network graph in $n$-dimensions or directions (that are mutually perpendicular) where $n$ is the number of vertices in the graph. The projection in each direction is represented in the form of a scalar value (called the eigenvalue) and its corresponding vector with entries for each vertex (called the eigenvector). Though the number of dimensions in the spectrum is the number of vertices in the graph, most of the variations could be captured in the first few dimensions of the coordinate system represented by the eigenvalues and the eigenvectors. The largest eigenvalue of the projection is called the principal eigenvalue and the corresponding eigenvector is called the principal eigenvector (could be determined by executing the power iteration algorithm [6] on the adjacency matrix of a network graph). The principal eigenvalue and its corresponding eigenvector capture maximum amount of variability in the data (in the case of a network graph, the data are the edges connecting the vertices). In this paper, we make use of the principal eigenvalue (also called the spectral radius) of the adjacency matrix of complex network graphs to analyze the variations in node degree and correlate with the coefficient of variation of node degree. We specifically use the spectral radius ratio for node degree as the basis to evaluate the evolution of two commonly studied complex network models, such as the scale-free networks and small-world networks. To the best of our knowledge, we could not come across any related work that uses spectral radius as the basis to analyze the evolution of scale-free networks and small-world networks.

The rest of the paper is organized as follows: In Section 2, we introduce the power iteration method that was used in this research to calculate the spectral radius of a network graph. Sections 3 and 4 validate our hypothesis on scale-free networks and small-world networks respectively. Section 7 concludes the paper.

## 2. POWER ITERATION METHOD TO CALCULATE SPECTRAL RADIUS

The power iteration method can be used to calculate the principal eigenvalue (i.e., spectral radius) and the corresponding principal eigenvector of a graph based on its adjacency matrix. The eigenvector $X_{i+1}$ of a network graph at the end of the $(i+1)^{th}$ iteration is given by: $X_{i+1} = AX_i / \|AX_i\|$, where $\|AX_i\|$ is the normalized value of the product of the adjacency matrix A of a given graph and the tentative eigenvector $X_i$ at the end of iteration $i$. The initial value of $X_i$ is [1, 1, ..., 1], a column vector of all 1s, where the number of 1s correspond to the number of vertices in the graph. We continue the iterations until the normalized value $\|AX_{i+1}\|$ converges to that of the normalized value $\|AX_i\|$. The value of the column vector $X_i$ at this juncture is declared the principal eigenvector of the network graph and the normalized value to which the iterations converge is the principal eigenvalue (i.e., the spectral radius). Figure 1 illustrates an example for computation of the spectral radius on a network graph. The converged normalized value of 2.21 is the spectral radius of the graph (denoted $\lambda_{sp}(G)$). If $k_{min}$, $k_{avg}$ and $k_{max}$ are the minimum, average and maximum node degrees, then, $k_{min} \leq k_{avg} \leq \lambda_{sp}(G) \leq k_{max}$ [7].

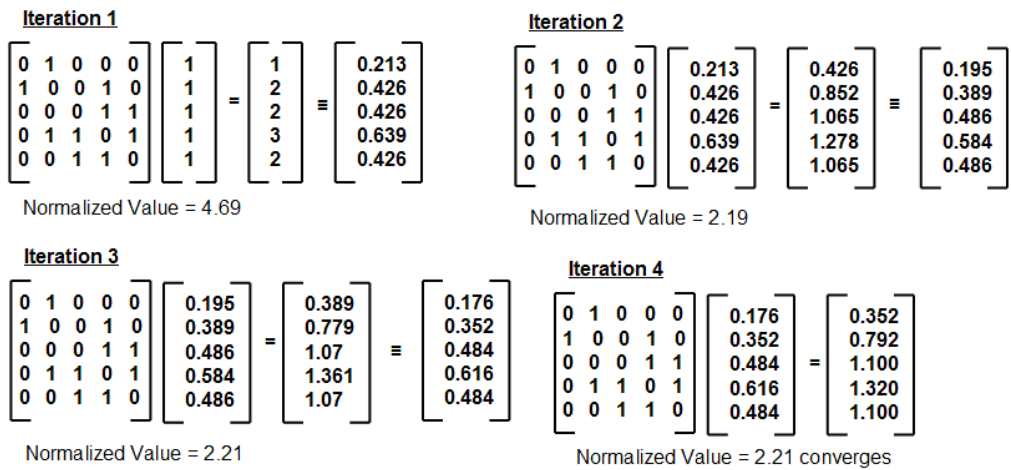

**Figure 1: Example to Illustrate the Execution of the Power Iteration Method to Compute the Spectral Radius of a Network Graph**

# 3. SIMULATION AND ANALYSIS OF SCALE-FREE NETWORKS

A scale-free network is a network wherein a majority of the nodes have a very small degree and very few nodes (but appreciable number of nodes) have a very high degree [4]. An airline network is typically a scale-free network with certain airports being used as high-degree hubs with connections to other hubs as well as connections to nearby low-degree nodes/airports. The simulations for scale-free networks were conducted as follows: We use the Barabasi-Albert (BA) model [8] for generating scale-free networks. According to this model, to start with, the network is assigned an initial set of nodes (3, in our simulations): the links between which are chosen arbitrarily, as long as there is one link per node (accordingly, we consider each node for link inclusion and connect the node to a randomly chosen node, other than itself, from the initial set of nodes). We then introduce nodes, one at a time. At each time step, a new node is introduced to the network with $m$ number of new links (varied from 2, 3, 4, 5, 10, 20, 30, 40, 50 and 90) connecting to the nodes that are already exist in the network (at most one new link per node). If the number of new links that could be added to the network is less than the number of nodes in the network, then the number of new links is set to the number of nodes in the network (in such cases, the newly introduced is basically connected to every node that already exist in the network). Once the number of new links added per node becomes less than the number of nodes that already exist in the network, then the node to pair with is chosen according to the probability formula that denotes the probability at which an already existing node $i$ with degree $k_i$ at time instant $t$ (corresponds to the introduction of node with ID $t$) acquires a link. We compute such probabilities for the nodes that were introduced prior to time instant $t$ (i.e., node IDs 1,..., $t$-1).

$$\prod(k_i^t) = \frac{k_i}{\sum_{j=1}^{t-1} k_j}$$

Note that the denominator in the probability formula sums only the degrees of the nodes that already exist in the network before the introduction of the new node and its links (i.e., the new node and its impact on the node degrees is not considered in the denominator portion of the formula). The values reported in Figures 2-5 and Tables 1-2 for the Spectral Radius Ratio for Node Degree (ratio of the spectral radius of the adjacency matrix and the average node degree) and the coefficient of variation of node degree (ratio of the standard deviation to the average node degree) are average values obtained from 100 runs of the simulation code for a particular condition (initial number nodes, total number of nodes and number of new links added per node introduction).

**Table 1: Spectral radius ratio for node degree vs. coefficient of variation of node degree after the introduction of the last node in the network
(Scale-free model for network evolution - initial # nodes: 3, total # nodes: 1000)**

|  | Number of new links added per node at the time of node introduction | | | | | | | | | |
|---|---|---|---|---|---|---|---|---|---|---|
|  | 2 | 3 | 4 | 5 | 10 | 20 | 30 | 40 | 50 | 90 |
| $\lambda_{sp}(G)/k_{avg}$ | 2.63 | 2.35 | 2.09 | 2.09 | 1.83 | 1.64 | 1.55 | 1.48 | 1.43 | 1.31 |
| $k_{SD}/k_{avg}$ | 1.31 | 1.19 | 1.07 | 1.08 | 0.92 | 0.79 | 0.73 | 0.69 | 0.64 | 0.54 |

**Table 2: Spectral radius ratio for node degree vs. coefficient of variation of node degree after the introduction of the last node in the network
(Scale-free model for network evolution - initial # nodes: 3, total # nodes: 100)**

|  | Number of new links added per node at the time of node introduction | | | | | | | | | |
|---|---|---|---|---|---|---|---|---|---|---|
|  | 2 | 3 | 4 | 5 | 10 | 20 | 30 | 40 | 50 | 90 |
| $\lambda_{sp}(G)/k_{avg}$ | 1.75 | 1.53 | 1.43 | 1.45 | 1.28 | 1.15 | 1.09 | 1.05 | 1.03 | 1.00 |
| $k_{SD}/k_{avg}$ | 0.97 | 0.80 | 0.69 | 0.69 | 0.55 | 0.39 | 0.31 | 0.23 | 0.17 | 0.02 |

The spectral radius ratio for node degree could be used as a measure to determine the appropriate number of new links per node (that could be added to the network upon introduction of a node into the network) to achieve a desired threshold level of variation in node degree. As we divide the spectral radius by the average node degree, the spectral radius ratio for node degree can be used as a measure to judge the variation of node degree across networks of any size. Tables 1 and 2 display the spectral radius ratio for node degree and coefficient of variation of node degree after the introduction of the last node in the network. As we can see, for a fixed value of the number of new links added per node and the initial number of nodes, the smaller the value of the total number of nodes to be introduced to the network, the lower the variation in node degree. As we increase the number of new links added per node introduction, the distribution of links is relatively more even across all the nodes, especially when the total number of nodes in the network is lower.

**Table 3: Correlation coefficient between spectral radius ratio for node degree vs. coefficient of variation of node degree as a function of time - with the introduction of nodes in the network (Scale-free model for network evolution - initial # nodes: 3)**

| Total # Nodes | Number of new links added per node at the time of node introduction | | | | | | | | |
|---|---|---|---|---|---|---|---|---|---|
| | 2 | 3 | 4 | 5 | 10 | 20 | 30 | 40 | 50 | 90 |
| 1000 | 0.974 | 0.987 | 0.986 | 0.969 | 0.978 | 0.976 | 0.974 | 0.972 | 0.969 | 0.967 |
| 100 | 0.981 | 0.989 | 0.987 | 0.973 | 0.958 | 0.942 | 0.866 | 0.796 | 0.734 | 0.934 |

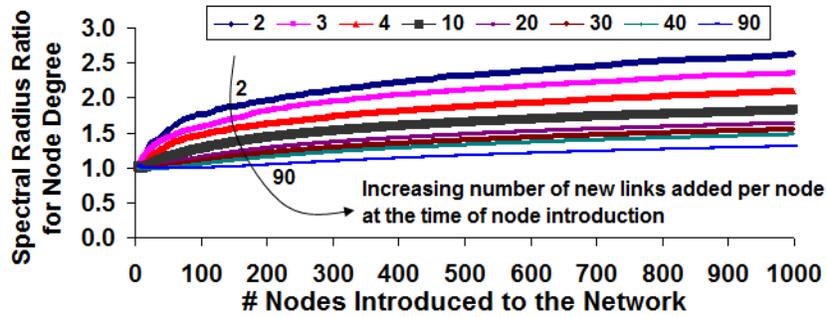

Figure 2: Distribution of the spectral radius ratio for node degree under the scale-free model for network evolution (initial # nodes: 3; total # nodes: 1000)

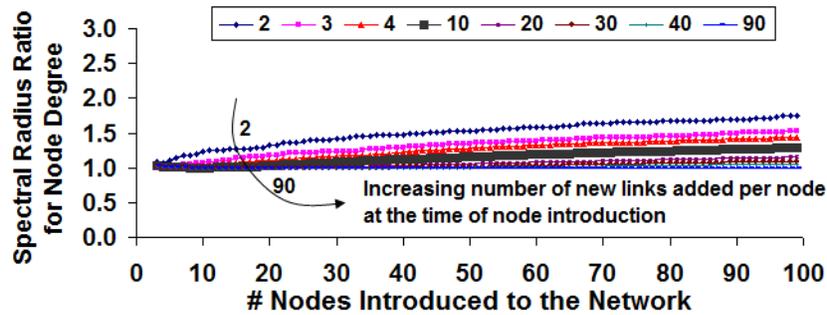

Figure 3: Distribution of the spectral radius ratio for node degree under the scale-free model for network evolution (initial # nodes: 3; total # nodes: 100)

Figures 2-3 and 4-5 respectively capture the distribution of the spectral radius ratio for node degree and the coefficient of variation of node degree as a function of time (upon the introduction of each new node into the network). After the initial set of nodes are assigned at least one link (per node), we add new nodes to the network, one at a time, as per the probability formula for an existing node to get a new link as a function of its degree. With passage of time, nodes that have been around in the network for a longer time incur a larger degree compared to nodes that were recently added; such a variation in node degree increases with time. Thus, we could see both the spectral radius ratio for node degree and the coefficient of variation of node degree to exhibit a very high correlation and similar pattern of distribution with time.

For a fixed value of the number of new links added per node and the initial number of nodes, the larger the value for the total number of nodes to be introduced to the network, the larger the space of distribution (range of values) for the two measures as a function of time as well as larger the correlation coefficient between the two measures. In other words, larger the scale-free nature of the network, the larger the correlation coefficient between the spectral radius ratio for node degree and the coefficient of variation of node degree. Table 3 displays the values for the correlation coefficient computed using the distribution of the spectral radius ratio for node degree and the coefficient of variation [9] of node degree as a function of time, with the introduction of new nodes into the network (for fixed values of the number of new links added per node introduction and the initial number of nodes). Apparently, when then the spectral radius ratio for node degree approaches close to 1 (scenario of total of 100 nodes with 90 new links per node introduction), the correlation coefficient between the two measures (that was henceforth decreasing with increase in the number of new links added per node) bounces back and approaches 1.

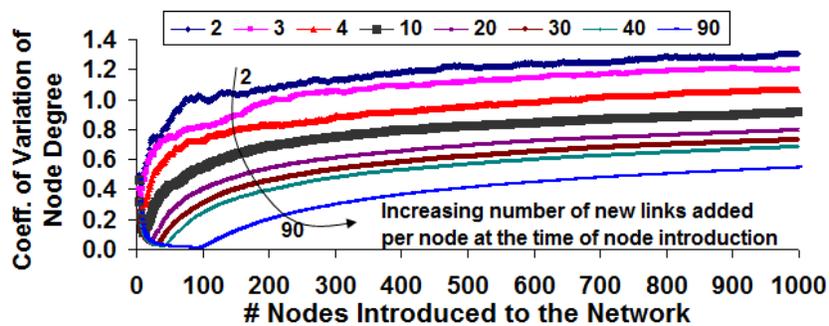

**Figure 4: Distribution of the coefficient of variation of node degree under the scale-free model for network evolution (initial # nodes: 3; total # nodes: 1000)**

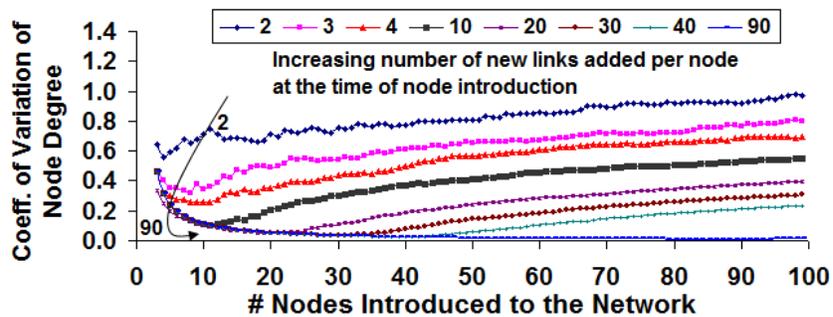

**Figure 5: Distribution of the coefficient of variation of node degree under the scale-free model for network evolution (initial # nodes: 3; total # nodes: 100)**

## 4. SIMULATION AND ANALYSIS OF SMALL-WORLD NETWORKS

A small-world network is a type of graph in which most nodes are not neighbors of one another, but most nodes can be reached from every other by a small number of hops [4]. An example for a small-world network is the road network of a country, wherein most of the cities are not neighbors of each other, but can be reached from one another by going through a fewer number of intermediate cities. The simulations for small-world networks are conducted according to the Watts and Strogatz model [10], explained as follows.

We start with a regular 1-dimensional ring network comprising of two concentric rings, each with the same number of nodes (say, $N$; hence, the total number of nodes in the network is referred to as $2N$). The nodes are initially organized in a ring network as shown in the left side of Figure 6. The number of neighbors per node is 4. For a node $i$ in the outer ring, its neighbors in the outer ring are chosen as follows: $(i + 1) \% N$ and $(i - 1 + N) \% N$. For a node $i$ in the inner ring, its neighbors in the inner ring are chosen as follows: $N + ((i + 1) \% N)$ and $N + ((i - 1 + N) \% N)$. In the initial regular ring network, as each node has 4 neighbors, the average number of neighbors per node is 4; with a total of $2N$ nodes, the total number of links in the initial regular ring network is (average node degree * total number of nodes / 2) = $(4 * 2N / 2) = 4N = 2*$total number of nodes in the network.

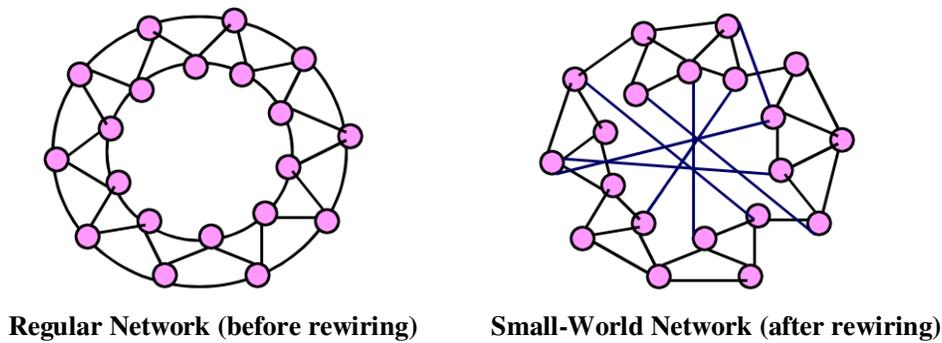

**Regular Network (before rewiring)**      **Small-World Network (after rewiring)**

**Figure 6: Transformation of a regular network to small-world network**

We consider each link in the initial set of links for a possible rewiring with a probability β. For each link $(u$-$v)$, where $u < v$ (in the order of the numerical IDs), we generate a random number (uniform-randomly distributed from 0 to 1) and if the random number is less than or equal to β, the node with the lower ID $u$ is paired with a randomly chosen node (other than itself, node $v$ and the current neighbors) and the newly created link is not considered for any rewiring. This way, the average number of neighbors per node at any time is still the initial value for the average number of neighbor per node (which is 4 in our simulations); but, variation in the distribution of the node degree starts creeping in with rewiring. The network at the right side of Figure 6 displays a sample network at the end of rewiring.

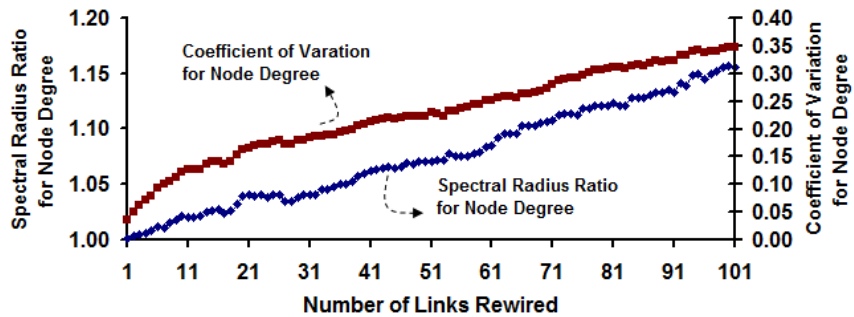

**Figure 7: Distribution of variation in node degree during the process of rewiring (total # nodes: 100; rewiring probability: 0.5)**

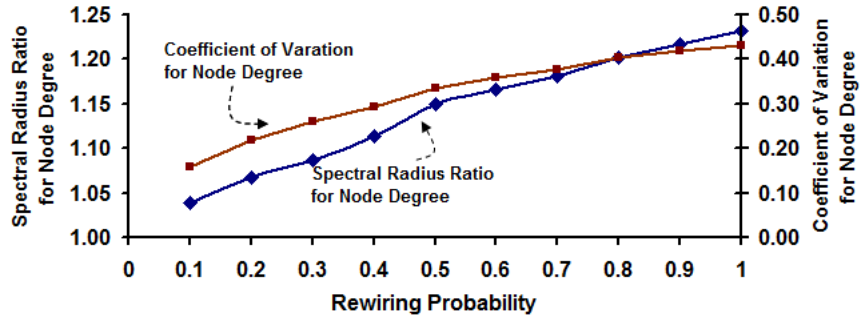

**Figure 8: Distribution of variation in node degree as a function of the rewiring probability (averaged across the total # nodes from 100 to 1000)**

We calculated the spectral radius of the network during the course of the rewiring of the links in the network (i.e., whenever a link is rewired). For a fixed total number of nodes and rewiring probability, as more links get rewired, the spectral radius of the network increased and the coefficient of variation in the node degree also increased (the correlation coefficient always remained above 0.95 during the process of rewiring). Figure 7 illustrates the distribution of the spectral radius for node degree and the coefficient of variation of node degree during the process of rewiring for a network with a total of 100 nodes (i.e., # nodes per ring $N$ is 50) and a rewiring probability of 0.5 - about 50% of the total number of links (0.5 * 2 * 100 = 100) got rewired and the figure displays the increase in the two metrics with rewiring.

For a fixed probability of rewiring, the variation in the node degree is about the same when the value for the total number of nodes in the network is varied (from 100 to 1000 in the simulations). In other words, for any fixed value of the rewiring probability, the spectral radius ratio for node degree observed at the end of rewiring a network of 100 nodes is about the same as the spectral radius ratio for node degree observed at the end of rewiring a network of 1000 nodes. This is a significant observation as the total number of nodes we start with does not impact the variation in the node degree during the course of rewiring (for a fixed probability of rewiring).

For a fixed value of the total number of nodes, smaller the value for the rewiring probability, smaller the variation in the node degree, as reflected in Figure 8 (averaged for each rewiring probability, based on the observations for the total number of nodes from 100 to 1000, as we do not see significant variation in node degree with increase in the number of nodes for a fixed probability of rewiring). We observe the spectral radius ratio for node degree to increase from 1 to 1.15 as we increase the rewiring probability from 0 to 0.5; but, as we increase the rewiring probability from 0.6 to 1 (the network being completely random), the spectral radius ratio for node degree increased only from 1.15 to 1.25.

## 5. CONCLUSIONS

The high-level contribution of this research is the identification of a positive correlation between the ratio of the spectral radius to the average node degree to that of the coefficient of variation of node degree for networks simulated from theoretical models (for scale-free networks and small-world networks). The positive correlation has been observed for networks of different sizes, ranging from networks of 5 nodes, to a few hundred nodes, all the way to networks of thousands of nodes. With this observation, we can now confidently say that the closer the value of the spectral radius ratio for node degree to 1.0, the smaller the variation among the degrees of the nodes in the network. The variation in the node degrees for any two networks can also be simply compared based on the spectral radius ratio for node degrees observed for the two networks, rather than requiring to compute the standard deviation of the node degrees for the two networks. As explained in Sections 3 and 4, the operating parameters of the theoretical models can be tuned to obtain a desired variation in the degree of the nodes in the network, measured in the form of the spectral radius ratio for node degree.